\documentclass[runningheads]{llncs}
\usepackage[misc]{ifsym}
\usepackage[T1]{fontenc}
\usepackage{pifont}
\usepackage{graphicx}
\usepackage{xcolor}
\usepackage{amsmath}
\usepackage{multirow}
\usepackage{amsfonts} 
\usepackage[utf8]{inputenc} 
\usepackage{bm} 
\usepackage{graphicx}
\usepackage{appendix}
\begin{document}
\title{Perspective$+$ Unet: Enhancing Segmentation with Bi-Path Fusion and Efficient Non-Local Attention for Superior Receptive Fields}
\titlerunning{Enhancing Segmentation with Bi-Path Fusion and Efficient Attention}
%
\author{Jintong Hu\inst{1} \and
Siyan Chen\inst{2} \and Zhiyi Pan\inst{1} \and Sen Zeng\inst{1} \and
Wenming Yang\inst{1(\textrm{\Letter})}}
\authorrunning{J. Hu et al.}
%
\institute{Shenzhen International Graduate School, Tsinghua University, Beijing, China \email{yang.wenming@sz.tsinghua.edu.cn}\\
\and
College of Electronics and Information Engineering, Shenzhen University, Shenzhen, China
\\
\url{https://github.com/tljxyys/Perspective-Unet}
}
\maketitle              
\begin{abstract}
Precise segmentation of medical images is fundamental for extracting critical clinical information, which plays a pivotal role in enhancing the accuracy of diagnoses, formulating effective treatment plans, and improving patient outcomes. Although Convolutional Neural Networks (CNNs) and non-local attention methods have achieved notable success in medical image segmentation, they either struggle to capture long-range spatial dependencies due to their reliance on local features, or face significant computational and feature integration challenges when attempting to address this issue with global attention mechanisms. To overcome existing limitations in medical image segmentation, we propose a novel architecture, Perspective$+$ Unet. This framework is characterized by three major innovations: (i) It introduces a dual-pathway strategy at the encoder stage that combines the outcomes of traditional and dilated convolutions. This not only maintains the local receptive field but also significantly expands it, enabling better comprehension of the global structure of images while retaining detail sensitivity. (ii) The framework incorporates an efficient non-local transformer block, named ENLTB, which utilizes kernel function approximation for effective long-range dependency capture with linear computational and spatial complexity. (iii) A Spatial Cross-Scale Integrator strategy is employed to merge global dependencies and local contextual cues across model stages, meticulously refining features from various levels to harmonize global and local information. Experimental results on the ACDC and Synapse datasets demonstrate the effectiveness of our proposed Perspective$+$ Unet. The code is available in the supplementary material.
\end{abstract}
\section{Introduction}

3D Medical image segmentation plays a crucial role in the realm of diagnostic radiology and surgical planning. Accurate segmentation of anatomical structures from volumetric datasets, such as magnetic resonance imaging (MRI) or CT, enables the extraction of critical clinical information, facilitating precise interventions, monitoring of disease progression and personalization of treatment strategies \cite{transunet,UnetR}. 

Recently, Convolutional Neural Networks (CNNs) have achieved tremendous success across various domains \cite{Binary,Meta-Learning,repvgg}. Their 3D-focused variants, like V-Net \cite{V-Net} and 3D U-Net \cite{Unet}, play a prominent role in local feature extraction for 3D medical image segmentation.
These models, including enhanced versions such as Res-UNet \cite{res-unet} with its residual connections, and U-Net$++$ \cite{Unet++} along with UNet3$+$ \cite{unet3+}, with their advanced skip pathways and nested designs, excel in detecting detailed local patterns. Despite their proficiency, the inherent limitation of CNNs' local receptive fields restricts their ability to capture broader spatial relationships essential for a comprehensive understanding of images. To compensate for this, multiple layers are required to expand the coverage, potentially leading to increased model complexity without effectively integrating global context.

Contrastingly, non-local methods expand the receptive fields of segmentation models, harnessing global context over mere local details in image analysis. These strategies focus on spatial relationships of the entire image to improve segmentation efficacy. Emblematics of this approach are transformer-based \cite{attention-unet,transunet} or hybrid transformer models \cite{scaleformer,missformer,eye-guide}, which apply global attention to image regions. Despite their objective to synthesize distant spatial information, these modules risk integrating noise and face quadratic computation costs tied to input size. While containing non-local interactions offers computational relief, it paradoxically trims crucial global insights. Efforts to balance computational efficiency and global context awareness in image segmentation have led to innovations like the Swin-Unet \cite{swin-unet,swin-tran} and Non-local Attention variants \cite{NLSA,efficient_nonlocal}, which partition images for blockwise attention calculation before global synthesis. Although this reduces complexity, it also limits the capture of wider spatial information.

In this paper, we propose Perspective$+$ Unet, a pioneering framework designed to augment the receptive field for enhanced 3D medical image segmentation. Our model incorporates an innovative Bi-Path encoder, utilizing two pathways: one utilizing standard convolutions to capture high-resolution local details, and the other employing dilated convolutions to capture broader contextual information. This configuration addresses issues commonly associated with dilated convolutions, allowing for the synthesis of a rich set of features with varying receptive field sizes, fostering a deeper and wider understanding of the imagery. Then, the extracted features are fed into an Efficient Non-Local Transformer Block (ENLTB), which builds upon the standard non-local attention mechanisms to provide a global perspective with a fraction of the usual computational overhead. The ENLTB effectively condenses long-range spatial dependencies into a compact feature set, enriching subsequent feature maps with globally attentive characteristics. In addition, to enhance the synergy between local and global feature representations, we design the Spatial Cross-Scale Integrator (SCSI) to encourage a coherent integration of information across different stages, ensuring that fine-grained details are preserved while benefiting from macroscopic contextual insights. The output of the SCSI module is incorporated into the decoder via skip connections, ultimately facilitating the generation of segmented images from the decoder output. Our model has been trained and tested on the ACDC and Synapse datasets, achieving competitive levels of performance in terms of both the Dice Similarity Coefficient (DSC) and Hausdorff Distance (HD) metrics. 

\begin{figure}[t]
\includegraphics[width=\textwidth]{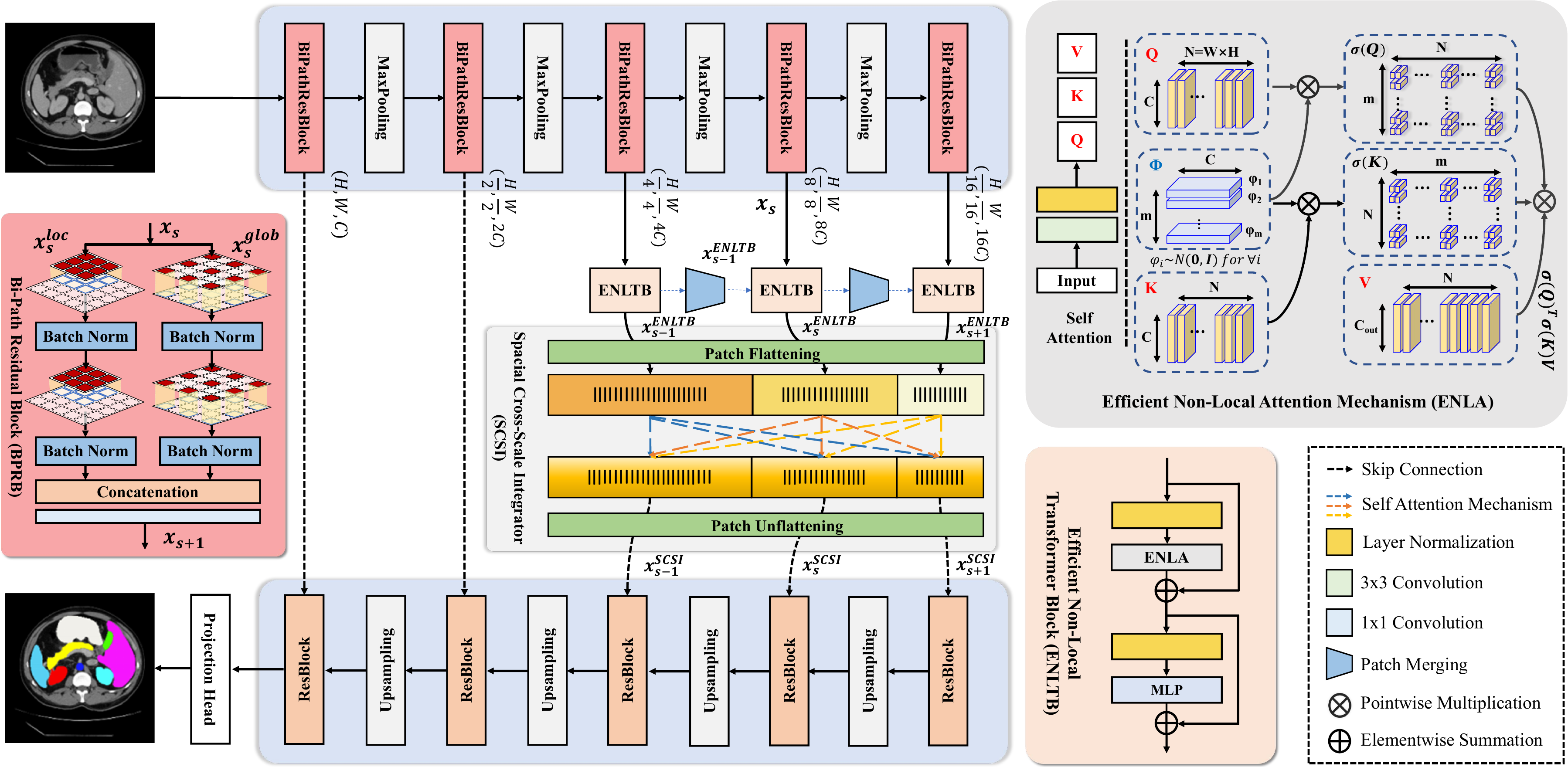}
\caption{The pipeline of the proposed Perspective$+$ Unet. The model consists of (i) a bi-path CNN based encoder to effectively capture local details and broad contextual information, (ii) a bottleneck composed of Efficient Non-Local Transformer Blocks and Spatial Cross-Scale Integrator for enhanced global perspective, and (iii) a decoder to incorporate both global and local information for generating segmentation results.} 
\label{fig1}
\end{figure}

\section{Methods}


Our proposed Perspective+ Unet adopts an encoder-bottleneck-decoder configuration, with the specific network workflow illustrated in Figure \ref{fig1}. In this chapter, we will provide a detailed introduction to the three modules within the network, starting with the Bi-Path Residual Block.

\subsection{Bi-Path Residual Block (BPRB)}

Increasing the size of the receptive field to improve accuracy has been a research focus in the advancement of 3D medical image segmentation. Traditionally, researchers sought to achieve this by stacking global modules, which, while expanding the view, often led to the loss of local region information: the broader receptive field to capture global information risked overlooking crucial local details.

To address this challenge, we have innovated the BPRB, a dual-pathway design that balances local and global information processing to enhance segmentation. One pathway uses dilated convolution to broaden the receptive field, understanding wider spatial information, but introducing feature discontinuity due to its spacing method. To solve this, BPRB incorporates another pathway with traditional convolution, focusing on capturing detailed features and maintaining local information continuity, thus achieving an optimal balance between global and local information processing. The proposed BPRB can formulated as:
\begin{equation}
    \boldsymbol{x}^{loc}_{s+1}=BatchNorm(f_s(BatchNorm(f_s(\boldsymbol{x}_{s}))))
\end{equation}
\begin{equation}
    \boldsymbol{x}^{glob}_{s+1}=BatchNorm(f^{k}_s(BatchNorm(f^{k}_s(\boldsymbol{x}_{s})))
\end{equation}
\begin{equation}
    \boldsymbol{x}_{s+1}=Conv(Concat[\boldsymbol{x}^{loc}_{s+1},\boldsymbol{x}^{glob}_{s+1}])
\end{equation}
where $\boldsymbol{x}^{loc}_{s+1}$, $\boldsymbol{x}^{glob}_{s+1}$, $\boldsymbol{x}_{s+1}$ are local, global, and final features in stage $s+1$. $f_s$ and $f^{k}_s$ are the convolution and dilated convolution with dilation rate $k$ in stage $s$.

\subsection{Efficient Non-Local Transformer Block (ENLTB)}

Capturing global information in input images is crucial for enhancing feature representation, which directly contributes to the effectiveness of segmentation. In response to this, we integrate the ENLTB as a strategic means to reconfigure the feature maps produced by the encoder, aiming to achieve enhanced contextual understanding and representational diversity. 

The ENLTB employs the Efficient Non-Local Self-Attention (ENLSA) mechanism, which substantially accelerates processing by substituting the conventional exponential kernel $\exp(\boldsymbol{Q}^{\top} \boldsymbol{K})$, referenced in non-local neural network \cite{Non-local}, with a more computationally efficient unbiased estimate $\sigma$. Specifically, $\exp(\boldsymbol{Q}^{\top} \boldsymbol{K})=\sigma(\boldsymbol{Q})^{\top}\sigma(\boldsymbol{K})$, where $\sigma(\cdot)$ can be reparameterized into a linear mapping. The detailed proofs are given below (constants unrelated to variables are omitted for enhanced readability):
\begin{align}
&\quad\ \exp \left(\boldsymbol{Q}^{\top} \boldsymbol{K}\right) \nonumber\\
&=\exp\left(\left\|\boldsymbol{Q}+\boldsymbol{K}\right\|^2/2\right) \cdot \exp\left(-\left(\left\|\boldsymbol{Q}\right\|^2+\left\|\boldsymbol{K}\right\|^2\right)/2\right) \nonumber\\
&= \exp\left(-\left(\left\|\boldsymbol{Q}\right\|^2+\left\|\boldsymbol{K}\right\|^2\right)/2\right)\int\exp\left(\left\|\boldsymbol{Q}+\boldsymbol{K}\right\|^2/2 -\left\|\boldsymbol{\phi}-\left(\boldsymbol{Q}+\boldsymbol{K}\right)\right\|^2/2\right) d \boldsymbol{\phi}\nonumber\\
& =\exp\left(-\left(\left\|\boldsymbol{Q}\right\|^2+\left\|\boldsymbol{K}\right\|^2\right)/2\right) \int \exp \left(-\|\boldsymbol{\phi}\|^2/2+\boldsymbol{\phi}^{\top}\left(\boldsymbol{Q}+\boldsymbol{K}\right)\right) d \boldsymbol{\phi}\nonumber\\
&=\mathbb{E}_{\boldsymbol{\phi} \sim \mathcal{N}\left(\mathbf{0}_c, \boldsymbol{I}_c\right)} \exp \left(\boldsymbol{\phi}^{\top}\left(\boldsymbol{Q}+\boldsymbol{K}\right)-\left(\left\|\boldsymbol{Q}\right\|^2+\left\|\boldsymbol{K}\right\|^2\right)/2\right) \nonumber\\
&=\mathbb{E}_{\boldsymbol{\phi} \sim \mathcal{N}\left(\mathbf{0}_c, \boldsymbol{I}_c\right)} 
\exp\left(-\left\|\boldsymbol{Q}-\boldsymbol{\phi}\right\|^2/2\right)\exp\left(-\left\|\boldsymbol{K}-\boldsymbol{\phi}\right\|^2/2\right)\nonumber\\
&=\sigma\left(\boldsymbol{Q}\right)^T \sigma\left(\boldsymbol{K}\right)
\end{align}
where $\boldsymbol{Q}$ and $\boldsymbol{K}$ stand for query and key in the attention. $\sigma(\boldsymbol{x})$ can be reparameterized as product of a multivariate normal variable $\boldsymbol{\Phi}$ and $\boldsymbol{x}$.

Relative to the traditional transformer's quadratic computational complexity, the ENLSA module only has a complexity of $O(2N)$ for the matrix projection computation of $\boldsymbol{Q}$ and $\boldsymbol{K}$, followed by $O(N)$ for the multiplication between $\sigma\left(\boldsymbol{Q}\right)^T \sigma\left(\boldsymbol{K}\right)$ and $\boldsymbol{V}$. This design ensures that the overall computational process is linearly related to the input size $N$, significantly reducing computational costs.

We have re-engineered the Transformer by replacing its attention mechanism with the ENLSA to construct the ENLTB. This modification streamlines feature representation enhancement. The first layer of ENLTB exclusively processes features received from the encoder. Subsequent ENLTB layers, however, aggregate both the features directly relayed from the encoder and those transmitted from the preceding ENLTB layer through patch merging, thereby intricately blending fine-grained details with coarse semantic information. The process of ENLTB can be represented as:

\begin{equation}
    \boldsymbol{x}_{s}^{in}=\begin{cases}
        \boldsymbol{x}_{s}, & \text{if } s=3\\
        concat[\boldsymbol{x}_{s}, PatchMerging(\boldsymbol{x}_{s-1}^{ENLTB})], & \text{else}
    \end{cases}
\end{equation}

\begin{align}
\boldsymbol{y}=\boldsymbol{x}_{s}^{in}+ENLSA\left ( LN\left ( \boldsymbol{x}_{s}^{in} \right )  \right ) , \quad 
\boldsymbol{x}_s^{ENLTB}=\boldsymbol{y} + MLP\left ( LN\left ( \boldsymbol{y} \right )  \right )
\end{align}
where the definition of $\boldsymbol{x}_{s}$ remains consistent with that presented in Section 2.1. $\boldsymbol{x}_s^{in}$, $\boldsymbol{x}_s^{ENLTB}$ denote the input and output of ENLTB at stage $s$. $LN$ and $MLP$ stand for the layer normalization and Multilayer Perceptron, respectively.

\subsection{Spatial Cross-Scale Integrator (SCSI) }
Collaborative feature interplay is instrumental in the enrichment of the network's interpretative performance, as a result, we design the SCSI module to ensure that detailed image complexities are segmented with both precision and finesse. 

The SCSI starts with patch flattening, where feature maps produced by every ENLTB are gathered and merged into a unified sequence. Following this, a transformer processes the sequence, which supports the learning of connections between different features. The refined feature sequences are then carefully mapped back onto their original feature maps for each scale, maintaining their original order of connection. 
The process of SCSI can be formulated as:
\begin{equation}
    \boldsymbol{x}_{conpound} = concat[flatten(\boldsymbol{x}_{s}^{ENLTB}),\cdots,flatten(\boldsymbol{x}_{s+t}^{ENLTB})
\end{equation}
\begin{equation}
    \boldsymbol{x}_{s}^{SCSI},\cdots,\boldsymbol{x}_{s+t}^{SCSI} = unflatten(transformer(\boldsymbol{x}_{conpound}))
\end{equation}
where $\boldsymbol{x}_{s}^{ENLTB}$ and $\boldsymbol{x}_{s}^{SCSI}$ denote the output of ENLTB and SCSI at stage $s$, respectively. $concat$ represents the concatenation operation.

The SCSI module equips the Perspective$+$ Unet with the capability to assimilate complementary information extending across scales, thereby simplifying resolution of the ambiguities linked with complex scale variations. 

\begin{table}[t]
\caption{Segmentation accuracy of different methods on the Synapse multi-organ CT
dataset. The best results are shown in \textbf{bold}.}
\label{table1}
\resizebox{\linewidth}{!}{
\begin{tabular}{c|c c|c c c c c c c c}
\hline
\multirow{2}{*}{Methods} & \multicolumn{2}{c|}{Average} & \multirow{2}{*}{Aorta} & \multirow{2}{*}{Gallbladder} & \multirow{2}{*}{Kidney(L)} & \multirow{2}{*}{Kidney(R)} & \multirow{2}{*}{Liver} & \multirow{2}{*}{Pancreas} & \multirow{2}{*}{Spleen} & \multirow{2}{*}{Stomach} \\ \cline{2-3}
~ & DSC $\uparrow$ & HD $\downarrow$ & ~ & ~ & ~ & ~ & ~ & ~ & ~ & ~ \\ \hline
U-Net \cite{Unet} & 76.85 & 39.70 & 89.07 & 69.72 & 77.77 & 68.60 & 93.43 & 53.98 & 86.67 & 75.58 \\
R50 Att-UNet \cite{transunet} & 75.57 & 36.97 & 55.92 & 63.91 & 79.20 & 72.71 & 93.56 & 49.37 & 87.19 & 74.95 \\
Att-UNet \cite{attention-unet} & 77.77 & 36.02 & 89.55 & 68.88 & 77.98 & 71.11 & 93.57 & 58.04 & 87.30 & 75.75 \\
R50 ViT \cite{transunet} & 71.29 & 32.87 & 73.73 & 55.13 & 75.80 & 72.20 & 91.51 & 45.99 & 81.99 & 73.95 \\
TransUnet \cite{transunet} & 77.48 & 31.69 & 87.23 & 63.13 & 81.87 & 77.02 & 94.08 & 55.86 & 85.08 & 75.62 \\
SwinUNet \cite{swin-unet} & 79.12 & 21.55 & 85.47 & 66.53 & 83.28 & 79.61 & 94.29 & 56.58 & 90.66 & 76.60 \\
AFTer-UNet \cite{afterunet} & 81.02 & - & \textbf{90.91} & 64.81 & 87.90 & 85.30 & 92.20 & 63.54 & 90.99 & 72.48 \\
ScaleFormer \cite{scaleformer} & 82.86 & 16.81 & 88.73 & \textbf{74.97} & 86.36 & 83.31 & 95.12 & 64.85 & 89.40 & 80.14 \\
MISSFormer \cite{missformer} & 81.96 & 18.20 & 86.99 & 68.65 & 85.21 & 82.00 & 94.41 & 65.67 & 91.92 & 80.81 \\
FCT \cite{FCT} & 83.53 & - & 89.85 & 72.73 & \textbf{88.45} & \textbf{86.60} & \textbf{95.62} & 66.25 & 89.77 & 79.42 \\
MSAANet \cite{MSAANet} & 82.85 & 18.54 & 89.40 & 73.20 & 84.31 & 78.53 & 95.10 & 68.85 & 91.60 & 81.78 \\
\hline
Perspective+ & \textbf{84.63} & \textbf{11.74} & 89.38 & 70.80 & 87.57 & 85.78 & 95.30 & \textbf{70.71} & \textbf{94.41} & \textbf{83.06} \\ \hline             
\end{tabular}
}
\end{table}

\begin{table}[b]
\caption{Segmentation accuracy of different methods on the ACDC dataset. The best results are shown in \textbf{bold}.}
\label{table2}
\centering
\setlength{\tabcolsep}{1mm}{
\begin{tabular}{c|c|c c c}
\hline
Methods & DSC $\uparrow$ & RV & Myo & LV \\ \hline
R50 U-Net \cite{transunet} & 87.55 & 87.10 & 80.63 & 94.92 \\
R50 Att-UNet \cite{transunet} & 86.75 & 87.58 & 79.20 & 93.47 \\
R50 ViT \cite{transunet} & 87.57 & 86.07 & 81.88 & 94.75 \\
TransUNet \cite{transunet} & 89.71 & 88.86 & 84.53 & 95.73 \\
SwinUNet \cite{swin-unet} & 90.00 & 88.55 & 85.62 & 95.83 \\
ScaleFormer \cite{scaleformer} & 90.17 & 87.33 & 88.16 & 95.04 \\
UNETR \cite{UnetR} & 88.61 & 85.29 & 86.52 & 94.02 \\
MCTE \cite{MCTE} & 91.31 & 89.14 & 89.51 & 95.27 \\
MISSFormer \cite{missformer} & 91.19 & 89.85 & 88.38 & 95.34 \\
nnFormer \cite{nnformer} & 92.06 & \textbf{90.94} & 89.58 & 95.65 \\
\hline
Perspective+ & \textbf{92.54} & 90.92 & \textbf{90.49} & \textbf{96.20} \\ \hline
\end{tabular}
}
\end{table}

\section{Experiments}

\subsection{Datasets and Evaluation}
\subsubsection{Synapse:}The Synapse dataset, from the MICCAI 2015 Multi-Atlas Abdomen Labeling Challenge, contains 30 abdominal 3D CT scans. Eighteen cases are randomly selected for training purposes, and the remaining 12 are allocated for testing. Our methodology is evaluated using average Dice Similarity Coefficient (DSC) and the average Hausdorff Distance (HD) for eight abdominal organs.

\subsubsection{ACDC:}The ACDC Challenge assembled MRI scan results from various patients. Each patient’s scan is manually annotated with three labels: Left Ventricle (LV), Right Ventricle (RV), and Myocardium (MYO). We randomly select 70 cases for training, 10 cases for validation, and 20 cases for testing. The methodology is evaluated using the average DSC.

\subsection{Experiment Details}

In our Perspective+ Unet, the encoder or decoder requires no pre-training and is entirely trained from scratch. To mitigate the risk of overfitting, a comprehensive suite of data augmentation techniques is employed, including probabilistic image flipping, additive Gaussian noise, Gaussian blur, contrast adjustment, and a variety of affine transformations such as scaling, rotation, shearing, and translation. Training is conducted on the Synapse and ACDC datasets, with a regimen of 600 and 1000 epochs, respectively, and a batch size of 12. Images are resized to 224$\times$224 pixels. To optimize the learning process, a learning rate of 0.05 is implemented and the weight decay is set at 0.0001 using the SGD optimizer. All training and testing procedures are performed on an NVIDIA RTX A5000 GPU.

\begin{figure}[t]
    \centering
    \includegraphics[width=1.0\linewidth]{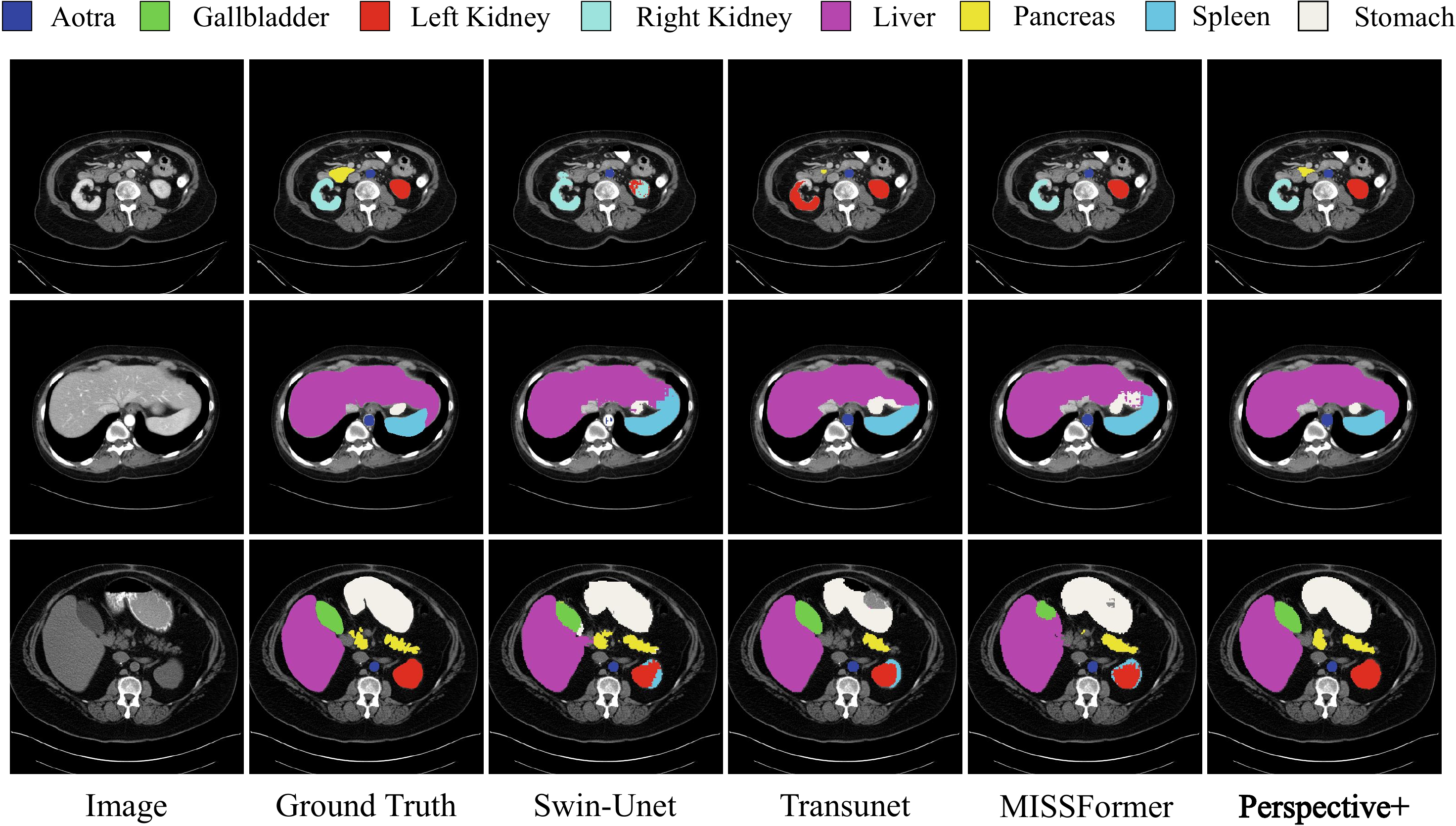}
    \caption{Visualized segmentation results of different methods on the Synapse multi-organ CT dataset. Our method (the last column) exhibits the smoothest boundaries and the most accurate segmentation outcomes.}
    \label{fig2}
\end{figure}

\subsection{Quantitative and qualitative segmentation results}
To validate the efficacy of Perspective+ Unet, we compare it against 11 advanced methods on the Synapse dataset. Highlighted in Table \ref{table1}, Perspective+ Unet excelled in DSC at 84.63$\%$ and HD at 11.74$\%$, showing marked improvements over the latest method, MSAANet, by 1.78$\%$ in DSC and a reduction in HD by 6.8mm. This advancement is credited to the BPRB and ENLTB modules, enhancing the model's receptive field. Further validation on the ACDC dataset confirmed Perspective+ Unet's superior versatility and robustness in capturing critical representations, as shown in Table \ref{table2}. The segmentation results of different methods on the Synapse dataset are shown in Figure \ref{fig2}. 

In our comparative analysis, ENLSA is evaluated against another non-local approach in terms of parameter count and number of floating point operations (FLOPs). The results presented in Table \ref{table3} demonstrate that ENLSA significantly outperforms NLSA in efficiency.

\begin{figure}[t]
    \centering
    \includegraphics[width=1.0\linewidth]{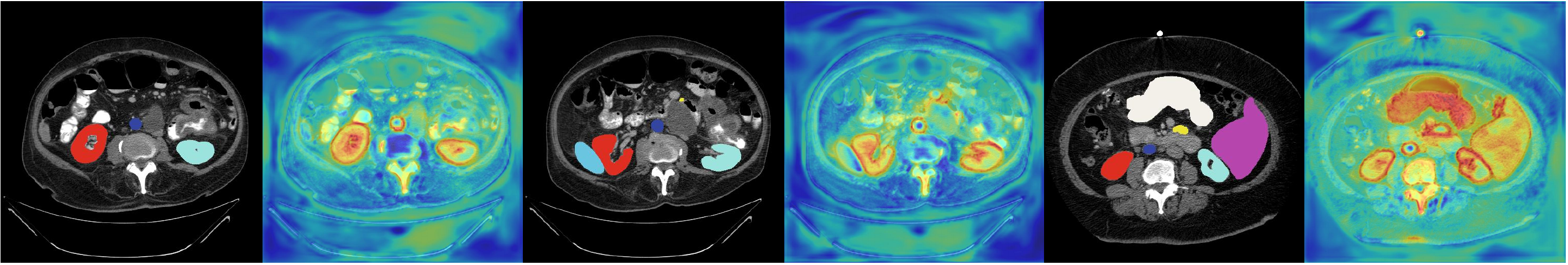}
    \caption{Visualization of attention heat maps from the intermediate layers of the network. Highlighting areas are closely aligned with segmentation labels, demonstrating our Perspective$+$ Unet's accuracy in feature identification and localization.}
    \label{fig3}
\end{figure}

Furthermore, to further validate the effectiveness of the model, this study employs visualization of the outputs from the intermediate layers of the network, as well as the creation of attention heat maps. As demonstrated in Figure \ref{fig3}, the highlighted areas within the heat map closely align with the segmentation labels. This not only proves the model's capability to efficiently identify key features from the input images, but also showcases its accuracy in feature localization.

\begin{table}
\begin{minipage}[htp]{0.4\textwidth}
\makeatletter\def\@captype{table}
\centering
\caption{Efficiency comparison of ENLSA. -m is the number of input channel. The input resolution is set as 224$\times$224.}
\label{table3}
\begin{tabular}{c|c|c}
\hline
Module & FLOPs$\downarrow$ & Params$\downarrow$ \\ \hline
NLSA-64 \cite{NLSA} & 2.06G & 0.04M \\
ENLSA-64 & \textbf{0.62G} & \textbf{0.01M} \\ \hline
NLSA-256 & 32.88G & 0.66M \\
ENLSA-256 & \textbf{9.87G} & \textbf{0.20M} \\
\hline
\end{tabular}
\end{minipage}
\hfill
\begin{minipage}[htp]{0.55\textwidth}
\makeatletter\def\@captype{table}
\centering
\caption{Ablation study on the impact of modules. BPRB: Bi-Path Residual Block, ENLTB: Efficient Non-Local Transformer Block, SCSI: Spatial Cross-Scale Integrator.}
\label{table4}
\begin{tabular}{c|c|c|c|c|c}
\hline
Model & BPRB & SCSI & ENLTB & DSC$\uparrow$ & HD$\downarrow$ \\ \hline
Setting 1 & \ding{56} & \ding{56} & \ding{56} & 84.04 & 16.63 \\
Setting 2 & \ding{52} & \ding{56} & \ding{56} & 83.36 & 14.70 \\
Setting 3 & \ding{52} & \ding{52} & \ding{56} & 83.92 & 13.94 \\
Setting 4 & \ding{52} & \ding{52} & \ding{52} & \textbf{84.63} & \textbf{11.74} \\ \hline
\end{tabular}
\end{minipage}
\end{table}

\subsection{Ablation Study}
In our proposed Perspective+ Unet architecture, we perform an in-depth ablation study on the added modules: BPRB, SCSI, and ENLTB, to assess their impact on model performance. Data from Table \ref{table4} reveal that the incorporation of the BPRB module alone slightly reduces DSC by -0.68$\%$, but significantly improves precision by reducing HD by 1.93mm. Further integration of the SCSI module with BPRB enhances both DSC by 0.56$\%$ and reduces HD by 0.76mm, evidencing the SCSI module's role in increasing reconstruction accuracy and precision. Employing all three modules achieves optimal outcomes, with a DSC of 84.63$\%$ and an HD of 11.74mm, highlighting each module's critical contribution to enhancing model performance.

\section{Conclusion}
In conclusion, Perspective+ Unet is distinguished by its innovative approach to enhancing spatial perception in 3D medical image segmentation. We design a dual-channel strategy that combines traditional and dilated convolutions with a novel transformer block and spatial cross-scale integrator. This architecture ensures comprehensive integration of image features across different stages and effectively handles both global structures and fine-grained parts, significantly enhancing the receptive field. Our experimental results demonstrate that Perspective+ Unet exhibits good segmentation accuracy in the ACDC and Synapse datasets, highlighting its significant potential in medical analysis.


%
%
%
\bibliographystyle{splncs04}
\bibliography{refs}
%




\newpage
\appendix

\section{Supplementary Material}

\begin{table}
\centering
\label{table5}
\vspace{0.5cm}
\caption{Detailed information of the two datasets we used. }
\begin{tabular}{c|c|c|c}
\hline
Datasets & Modality & Num. of Class & Train\textbackslash Valid\textbackslash Test \\ \hline
Synapse dataset & CT & 9 & 18\textbackslash0\textbackslash12 \\
ACDC dataset & MRI & 4 & 70\textbackslash10\textbackslash 20 \\
\hline
\end{tabular}
\end{table}

\begin{table}

\centering
\label{table6}
\caption{Efficiency comparison of ENLSA. -m is the number of input channel.}
\begin{tabular}{c|c|c|c}
\hline
Module & Input size & FLOPs$\downarrow$ & Params$\downarrow$ \\ \hline
NLSA-64 & [1, 64, 224, 224] & 2.06G & 0.04M \\
ENLSA-64 & [1, 64, 224, 224] & \textbf{0.62G} & \textbf{0.01M} \\ 
NLSA-128 & [1, 128, 112, 112] & 2.06G & 0.16M \\
ENLSA-128 & [1, 128, 112, 112] & \textbf{0.62G} & \textbf{0.05M} \\ 
NLSA-256 & [1, 256, 56, 56] & 2.06G & 0.66M \\
ENLSA-256 & [1, 256, 56, 56] & \textbf{0.62G} & \textbf{0.20M} \\ 
NLSA-512 & [1, 512, 28, 28] & 2.06G & 2.62M \\
ENLSA-512 & [1, 512, 28, 28] & \textbf{0.62G} & \textbf{0.79M} \\ 
NLSA-1024 & [1, 1024, 14, 14] & 2.06G & 10.49M \\
ENLSA-1024 & [1, 1024, 14, 14] & \textbf{0.62G} & \textbf{3.15M} \\

\hline
\end{tabular}
\end{table}

\begin{table}

\centering
\label{table7}
\caption{Inputs and outputs size for each stage of the BPRB and ENLTB modules.}
\begin{tabular}{c|c|c|c|c}
\hline
\multirow{2}{*}{Stage} & \multicolumn{2}{c|}{BPRB} & \multicolumn{2}{c}{ENLTB} \\ \cline{2-5}
~ & Input size & Output size & Input size & Output size \\ \hline
1 & [1, 3, 224, 224] & [1, 64, 224, 224] & - & - \\
2 & [1, 64, 224, 224] & [1, 128, 112, 112] & - & - \\
3 & [1, 128, 112, 112] & [1, 256, 56, 56] & [1, 256, 56, 56] & [1, 256, 56, 56] \\
4 & [1, 256, 56, 56] & [1, 512, 28, 28] & [1, 512, 28, 28] & [1, 512, 28, 28] \\ 
5 & [1, 512, 28, 28] & [1, 1024, 14, 14] & [1, 1024, 14, 14] & [1, 1024, 14, 14] \\ \hline
\end{tabular}
\end{table}

\begin{figure}[t]
    \centering
    \includegraphics[width=1.0\linewidth]{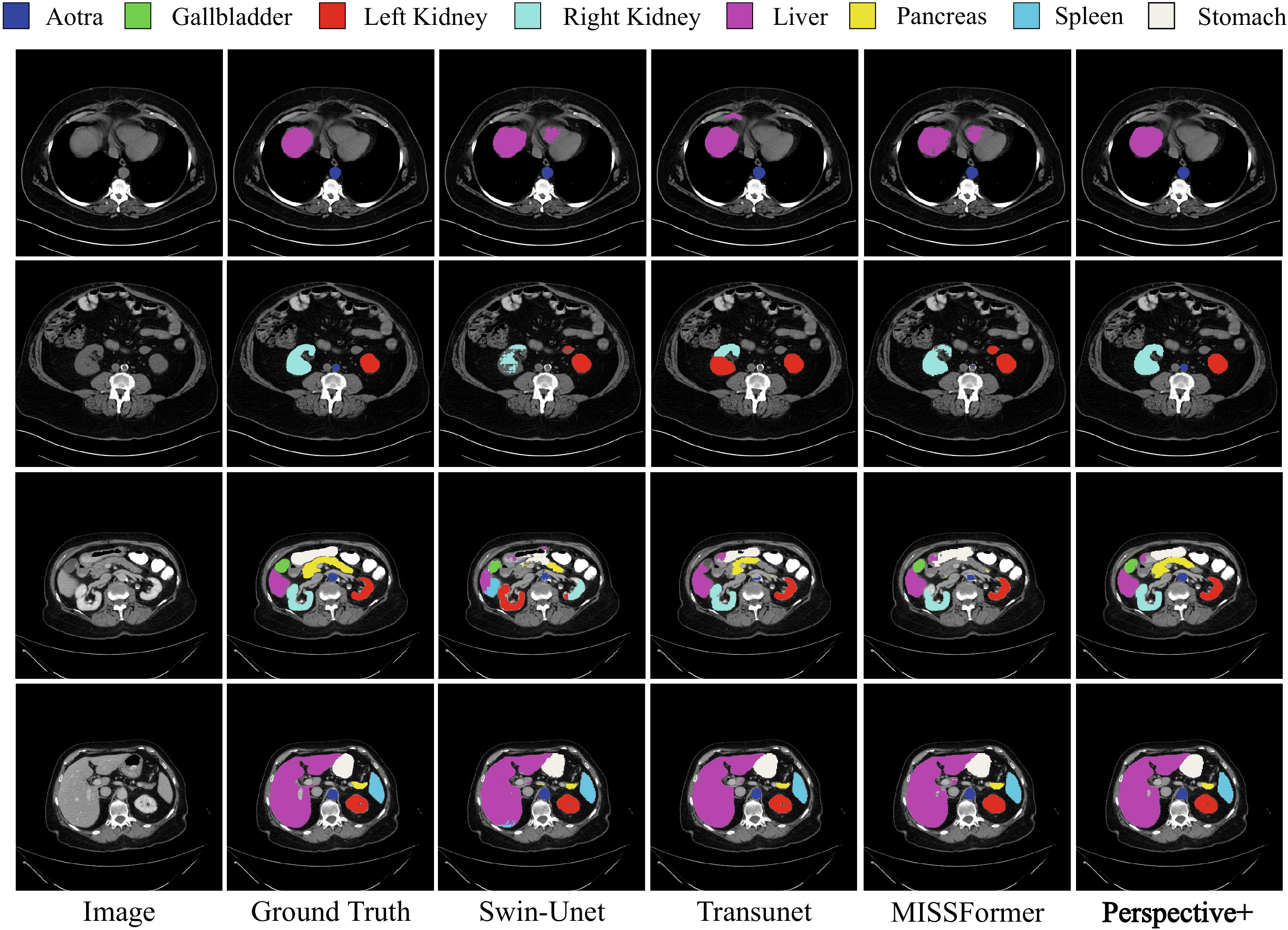}
    \caption{Visualized segmentation results of different methods on the Synapse multi-organ CT dataset. Our method (the last column) exhibits the smoothest boundaries and the most accurate segmentation outcomes.}
    \label{fig4}
\end{figure}

\begin{figure}[t]
    \centering
    \includegraphics[width=1.0\linewidth]{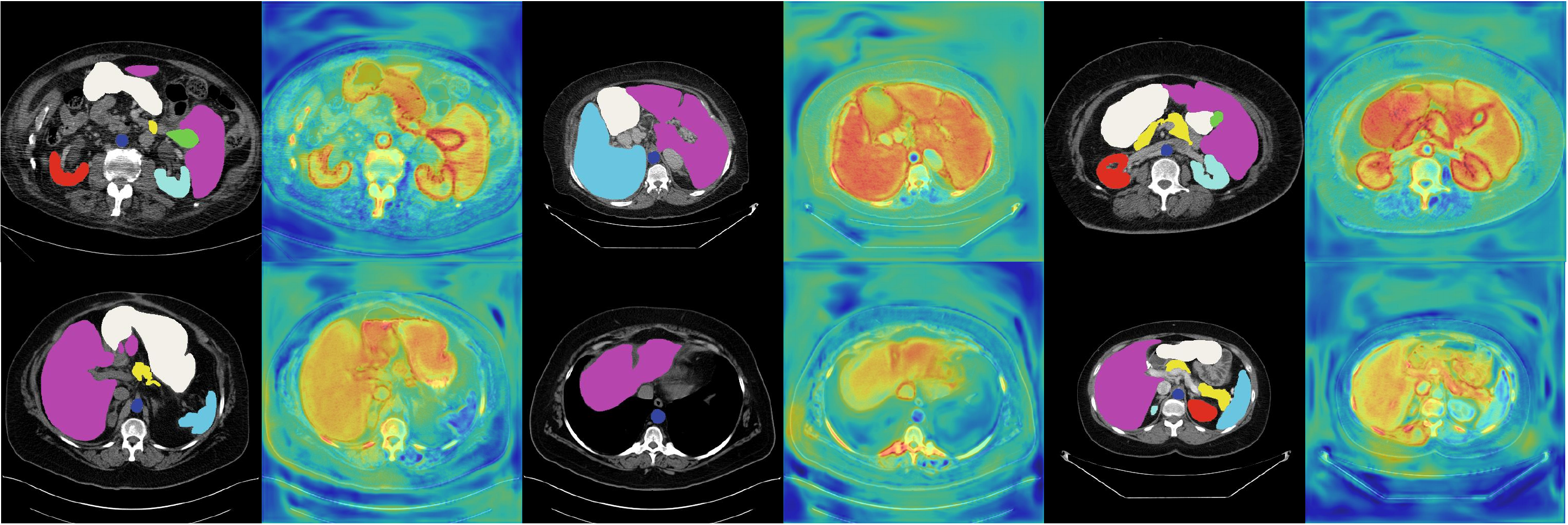}
    \caption{Visualization of attention heat maps from the intermediate layers of the network. Highlighting areas are closely aligned with segmentation labels, demonstrating our Perspective$+$ Unet's accuracy in feature identification and localization.}
    \label{fig5}
\end{figure}

\end{document}